\documentclass[amsmath,amssymb,amsfonts,twocolumn]{revtex4}
\usepackage{graphicx}
\usepackage{bbm}
\usepackage{float}
\usepackage{braket,bigints}
\usepackage{amsmath,stmaryrd,mathtools}
\def \beq {\begin{equation}}
\def \eeq {\end{equation}}
\def \tr {\rm Tr}
\newcommand {\qs} {{\rm Q}_{\rm S}}
\newcommand {\qt} {{\rm Q}_{\rm T}}
\newcommand {\ks} {k_{\rm S}}
\newcommand {\kt} {k_{\rm T}}

\begin{document}
\title{Quantum relative entropy shows singlet-triplet coherence is a resource in the radical-pair mechanism of biological magnetic sensing} 
\author{I. K. Kominis}
\affiliation{Department of Physics, University of Crete, 71003 Heraklion, Greece\\ Institute for Theoretical and Computational Physics, University of Crete, 70013 Heraklion, Greece}

\begin{abstract}
Radical-pair reactions pertinent to biological magnetic field sensing are an ideal system for demonstrating the paradigm of quantum biology, the exploration of quantum coherene effects in complex biological systems. We here provide yet another fundamental connection between this biochemical spin system and quantum information science. We introduce and explore a formal measure quantifying singlet-triplet coherence of radical-pairs using the concept of quantum relative entropy. The ability to quantify singlet-triplet coherence opens up a number of possibilities in the study of magnetic sensing with radical-pairs. We first  use the explicit quantification of singlet-triplet coherence to affirmatively address the major premise of quantum biology, namely that quantum coherence provides an operational advantage to magnetoreception. Secondly, we use the concept of incoherent operations to show that incoherent manipulations of nuclear spins can have a dire effect on singlet-triplet coherence when the radical-pair exhibits electronic-nuclear entanglement. Finally, we unravel subtle effects related to exchange interactions and their role in promoting quantum coherence. 
\end{abstract}
\maketitle
\section{Introduction}
Radical-pair reactions \cite{steiner}, central in understanding avian magnetoreception \cite{ritz,johnsen} and spin transport in photosynthetic reaction centers \cite{matysik_review}, have in recent years become a flourishing paradigm of quantum biology \cite{plenio_review,kominis_review}, the study of quantum coherence effects in complex biological systems \cite{Engel,Caruso,Briegel,Kais,Guo}. 
Indeed, it was shown that the understanding of the fundamental quantum dynamics of radical-pair reactions rests on quantum meaurement theory, quantum coherence quantifiers and quantum trajectories elucidating the physics at the single molecule level \cite{PRE2009,PRE2011,PRE2014,CPL2015}. These works also led to a new master equation describing radical-pair spin dynamics \cite{kominis_review}, qualitatively and quantitatively departing from the theory \cite{Haberkorn} traditionally used in spin chemical calculations. Moreover, quantum information concepts like violation of entropy bounds were used to further demonstrate \cite{PRE2017} the inadequacy of the long-standing theoretical foundations of spin chemistry \cite{HoreReview} in a general way unaffected by the precise knowledge, or lack thereof, of molecular parameters. Most recently, a quantum metrology approach addressed the fundamental limits to quantum sensing of magnetic fields using radical-pair reactions by treating them as biochemical quantum magnetometers \cite{PRA2017}.

Following these developments, there have been several other approaches exploring radical-pair quantum dynamics \cite{JH, ManolopoulosJCP2018,Hore2019}, essentially concurring with the basic aforementioned findings, namely that a new fundamental theory based on quantum measurements is required to understand these spin-dependent biochemical reactions, that tools of modern quantum metrology are indeed useful to address their dynamics, and that in general, radical-pairs are an ideal system demonstrating the paradigm of quantum biology.

Yet apart from any quantitative or qualitative differences in the various approaches being explored, it is broadly accepted that radical-pairs {\it do} exhibit quantum coherence, in particular {\it singlet-triplet coherence} defined by the pair's two electronic spins. The role of global quantum coherence in magnetoreception has been addressed \cite{PlenioPRL2013}, but the role of {\it singlet-triplet coherence} has not been explored, because apart from some empirical approaches \cite{PRE2014}, a formal and physically intuitive measure of singlet-triplet coherence has been lacking. Moreover, there have been discussions on whether a semiclassical treatment could replace coherent spin dynamics in radical-pair magnetoreception \cite{HoreFD2020}, but again, such discussions are phenomenological and of limited predictive power unless a concrete singlet-triplet coherence measure is established. This way it will be straightforward to understand the classical limit of the relevant dynamics by gradually eliminating singlet-triplet coherence, {\it while being able to exaclty quantify its presence}. 

Interestingly, during  the last few years quantum coherence measures have been formally investigated in quantum information science in the context of resource theories \cite{PlenioPRL2014,PlenioRMP,QCQ1,QCQ2,QCQ3,QCQ4,QCQ5}. In this work we introduce and formally analyze quantum relative entropy as a singlet-triplet coherence measure. In particular, we formulate a quantum relative entropy measure for quantifying coherence {\it between two subspaces} of the total Hilbert space, which should be generally useful irrespective of the particular physical system under consideration herein.

As the main application of our singlet-triplet quantifier, we {\it affirmatively, concretely and quantitatively} address the {\it main premise of quantum biology}, namely that singlet-triplet coherence indeed provides for an operational advantage in magnetoreception. We investigate the correlation of singlet-triplet coherence with the figure of merit for the operation of radical-pair reactions as a compass. Having an explicit quantifier of singlet-triplet coherence, we can controllably suppress it while always quantifying it, and thus study the compass figure of merit as we transition from a highly coherent to a highly incoherent regime. 

As a byproduct of introducing the so-called incoherent operations while setting up the formal singlet-triplet coherence quantification, we show that a class of incoherent operations originate from operating on just the nuclear spins of the radical-pair. This can have dire consequences for the electronic singlet-triplet coherence when there is electron-nuclear entanglement, a finding opening up a number of studies on the effect of nuclear spin dynamics on the radical-pair spin dynamics.  

As another application, we study the role of specific magnetic interactions in promoting singlet-triplet coherence, in particular the exchange interaction, and find subtle effects alluding to optimal values of exchange couplings promoting the quantum advantage of singlet-triplet coherence.

The structure of the paper is the following. In Sec. II we provide the basic definitions, motivate the need to introduce a measure of singlet-triplet cohererence in simple and intuitive terms, and recapitulate our previous empirical attemps at defining such a measure. In Sec. III we present the definition of a singlet-triplet coherence quantifier based on relative entropy and fully analyze its properties at a formal level. In Sec. IV we study magnetoreception and argue quantitatively in support of the main premise of quantum biology. We conclude with a summary and an outlook in Sec. V.
\section{Definitions, motivation and previous work}
\begin{figure}
\begin{center}
\includegraphics[width=7. cm]{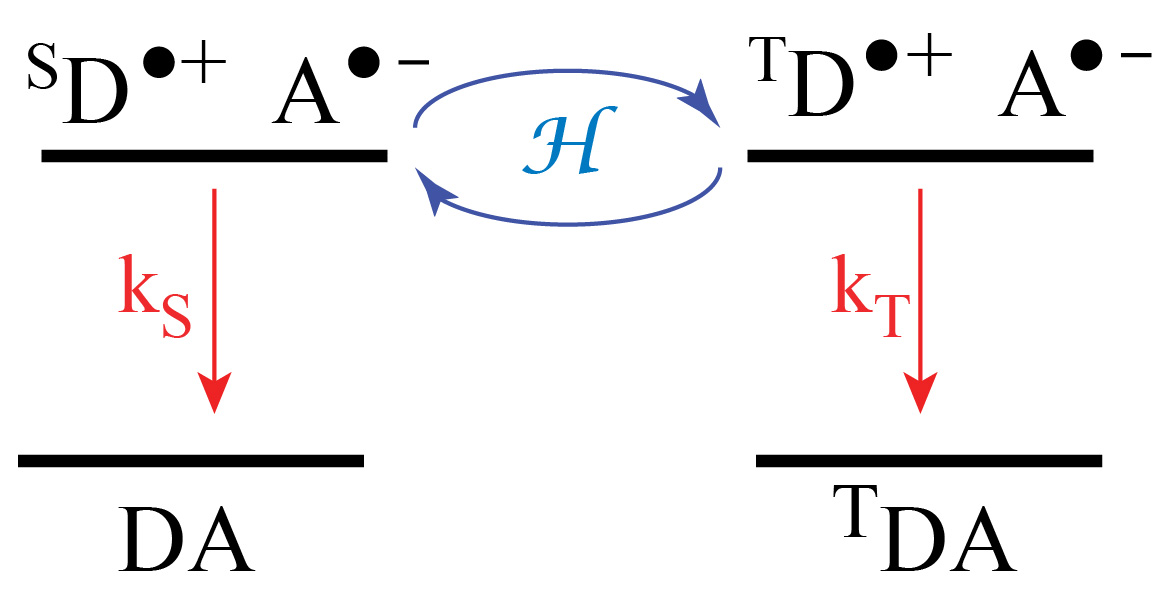}
\caption{Radical-pair reaction dynamics. A charge transfer following the photoexcitation (not shown here) of a donor-acceptor dyad DA produces a singlet state radical-pair $^{\rm S}{\rm D}^{\bullet +}{\rm A}^{\bullet -}$, which is coherently converted to the triplet radical-pair, $^{\rm T}{\rm D}^{\bullet +}{\rm A}^{\bullet -}$, due to intramolecule magnetic interactions embodied in the spin Hamiltonian ${\cal H}$. Simultaneously, spin-selective charge recombination leads to singlet and triplet neutral products with respective rates $\ks$ and $\kt$.}
\label{schematic}
\end{center}
\end{figure}

Radical-pair reaction dynamics are depicted in Fig. \ref{schematic}. A charge transfer following the photoexcitation (not shown) of a donor-acceptor dyad DA leads to the radical-pair ${\rm D}^{\bullet +}{\rm A}^{\bullet -}$, where the two dots represent the two unpaired electron spins of the two radicals. The initial state of the two unpaired electrons of the radical-pair is usually a singlet, denoted by $^{\rm S}{\rm D}^{\bullet +}{\rm A}^{\bullet -}$. Both D and A contain a number of magnetic nuclei (their initial state is usually fully mixed) which hyperfine-couple to the respective unpaired electron of D and A. The resulting magnetic Hamiltonian ${\cal H}$ involves all such hyperfine couplings, and extra terms accounting for the electronic Zeeman interaction with an applied magnetic field, exchange and dipolar interactions etc. 

The initial electron singlet state (and for that matter the triplet-state) is not an eigenstate of ${\cal H}$, hence the initial formation of $^{\rm S}{\rm D}^{\bullet +}{\rm A}^{\bullet -}$ is followed by singlet-triplet (S-T) mixing, a {\it coherent} oscillation of the spin state of the electrons (and concomitantly the nuclear spins), designated by $^{\rm S}{\rm D}^{\bullet +}{\rm A}^{\bullet -}\leftrightharpoons~^{\rm T}{\rm D}^{\bullet +}{\rm A}^{\bullet -}$. This coherent spin motion has a finite lifetime since charge recombination, i.e. charge transfer from A back to D, terminates the reaction and leads to the formation of either singlet or triplet neutral reaction products, produced at the rate $\ks$ and $\kt$, respectively. Both rates are in principle known or measurable parameters (in general different) of the specific molecular system under consideration.  

When discussing radical-pair states, we refer to a Hilbert space comprising of two electron spins, one for each radical, and any number of nuclear spins residing in both radicals. For the simplest possible approach, one could even consider a fictitious radical-pair without nuclear spins. In such a case, the S-T basis states of the two-electron system are denoted by $\ket{\rm s}=(\ket{\uparrow\downarrow}-\ket{\downarrow\uparrow})/\sqrt{2}$ for the singlet, and $\ket{{\rm t}_j}$ for the triplets ($j=0,\pm 1$), where $\ket{{\rm t}_1}=\ket{\uparrow\uparrow}$, $\ket{{\rm t}_0}=(\ket{\uparrow\downarrow}+\ket{\downarrow\uparrow})/\sqrt{2}$ and $\ket{{\rm t}_{-1}}=\ket{\downarrow\downarrow}$. When considering radical-pairs with nuclear spins, the tensor 
product structure will be explicitly given. For example, for the case of a single nuclear spin-1/2, a singlet electonic state and a spin-up nucleus will be denoted as $\ket{\rm s}\otimes\ket{\Uparrow}$.
\subsection{Singlet and triplet projectors}
The singlet and triplet projection operators are $\qs=(1/4-\mathbf{s}_1\cdot\mathbf{s}_2)\otimes\mathbbmtt{1}$ and $\qt=(3/4+\mathbf{s}_1\cdot\mathbf{s}_2)\otimes\mathbbmtt{1}$, with $\mathbbmtt{1}$ being the unit operator in the nuclear spin space (the dimension of the unit matrix should be evident from the context), and $\mathbf{s}_1$ and $\mathbf{s}_2$ the electron spins of the two radicals.  The projectors $\qs$ and $\qt$ are orthogonal and complete, i.e.  $\qs\qt=\qt\qs=0$ and $\qs+\qt=\mathbbmtt{1}$ (here $\mathbbmtt{1}$ refers to the total Hilbert space electrons+nuclei). 

Using the completeness property we can multiply any given radical-pair density matrix $\rho$ from left and right with $\mathbbmtt{1}=\qs+\qt$ and write 
\beq
\rho=\rho_{SS}+\rho_{TT}+\rho_{ST}+\rho_{TS}\label{ident}
\eeq
where $\rho_{xy}={\rm Q}_{x}\rho {\rm Q}_{y}$, with $x,y={\rm S,T}$. We will make frequent use of the identity \eqref{ident} in the following. We can already identify $\rho_{SS}+\rho_{TT}$ as the S-T "incoherent part" of $\rho$, and $\rho_{ST}+\rho_{TS}$ the S-T "coherent part" of $\rho$, to be formally defined and quantified later.

The density matrix, the projectors and any other operator relevant to a particular radical-pair having $M$ nuclei with nuclear spins $I_1$, $I_2$, ..., $I_M$ have dimension $d=4d_{\rm nuc}$, where 4 is the multiplicity of the two electron-spin space, and $d_{\rm nuc}=(2I_1+1)(2I_2+1)...(2I_M+1)$ is the dimension of the nuclear spin space.
\subsection{Motivation}
To motivate this work, we can ask a few questions, the answers to which are anything but obvious: {\bf (a)} Which state is "more" S-T coherent, $\ket{\psi_1}=(\ket{\rm s}+\ket{{\rm t}_0})/\sqrt{2}$, or $\ket{\psi_2}=(\ket{\rm s}+\ket{{\rm t}_1}+\ket{{\rm t}_0}+\ket{{\rm t}_{-1}})/2$ ? Clearly, both states are pure and normalized, and both involve a superposition of the singlet and some of the triplet states, but which is "{\it more coherent}"? Bringing nuclear spins into the picture, we might again ask: {\bf (b)} which state is "more" S-T coherent, $\ket{\psi_3}=(\ket{\rm s}\otimes\ket{\Uparrow}+\ket{{\rm t}_1}\otimes\ket{\Downarrow})/\sqrt{2}$ or $\ket{\psi_4}=(\ket{\rm s}\otimes\ket{\Uparrow}+\ket{{\rm t}_0}\otimes\ket{\Uparrow}+\ket{\rm t_1}\otimes\ket{\Uparrow})/\sqrt{3}$? Finally, regarding the magnetic interactions that directly drive or indirectly affect coherent S-T oscillations in the radical-pair state, which are at the heart of spin dynamics of this system, one might also ask: {\bf (c)} which are the interactions promoting S-T coherence and why ? 
\subsection{Previous measures of S-T coherence}
We first introduced the quantification of S-T coherence in radical-pairs in \cite{PRE2011}. The motivation was the fundamental description of reaction super-operators, which we claim depends on "how much" S-T coherent are the radical-pairs at any given time (see \cite{kominis_review} for further details). However, the current work is decoupled from this discussion, because here we choose $\ks=\kt=k$, simplifying the reaction dynamics considerably. For completeness, we list the previous S-T coherence quantifiers and comment on their shortcomings.

Our first quantifier \cite{PRE2011} was introduced in analogy with the coherence for a light field in a double-slit interferometer, 
$p_{\rm coh}={{\tr\{\rho_{ST}\rho_{TS}\}}/[{\tr\{\rho_{SS}\}\tr\{\rho_{TT}\}}}]$. One problem of this definition is that {\it any} coherent superposition e.g. $\alpha\ket{\rm s}+\beta\ket{\rm t_0}$, whatever the coefficients $\alpha$ and $\beta$, is mapped into a maximum (equal to 1) coherence, whereas it would make intuitive sense that the more asymmetrical the superposition (e.g. the closer to one is the singlet probability) the smaller the S-T coherence should be. Yet another problem is that $p_{\rm coh}$ is not a permissible measure based on the formal requirements set forth at \cite{PlenioPRL2014}, because it is a so-called $l_2$-norm, i.e. $p_{\rm coh}$ scales {\it with the square} of the off-diagonal elements of $\rho$. 

In \cite{PRE2014} we introduced the $l_1$-norm ${\cal C}\llbracket\rho\rrbracket={4\over 3}\sum_{j=0,\pm}\sqrt{\tr\{\rho_{\rm ST}|T_j\rangle\langle T_j|\rho_{\rm TS}\}}$.
This has another two shortcomings. First, the normalizing factor of 4/3 is incorrect, because the maximum value of the sum $\sum_{j=0,\pm}|\alpha_{\rm s}\beta_{j}|$ is not 3/4 but $\sqrt{3}/2$, and occurs for $|\alpha_{\rm s}|=1/\sqrt{2}$ and $|\beta_{j}|=1/\sqrt{6}$. Second, the sum $\sum_{j=0,\pm}|\alpha_{\rm s}\beta_{j}|$ is biased towards triplet states, i.e. it produces a higher coherence measure when more triplet states enter the superposition, {\it for a given triplet character of the state}. For example the superpositions ${1\over \sqrt{2}}\ket{\rm s}+{1\over\sqrt{6}}\sum_{j=-1}^1\ket{\rm t_j}$ and ${1\over\sqrt{2}}(\ket{\rm s}+\ket{\rm t_0})$ both have $\langle\qs\rangle=1/2$, i.e. for both it is equally uncertain whether they are found in the singlet or triplet subspace upon a measurement of $\qs$, yet the former is maximally coherent whereas the latter is not. However, we expect that maximum S-T coherence is attributed to all pure states having maximum quantum uncertainty in a measurement of $\qs$ or $\qt$. All aforementioned shortcomings are alleviated by the quantifier defined in the following based on the quantum relative entropy.
\section{Formal definition of S-T coherence in radical-pairs based on the quantum relative entropy}
We will now develop of formal theory of S-T coherence in radical-pairs using a central concept of quantum information, quantum relative entropy, adapted to our case. The resource theory of quantum cohence is built \cite{PlenioPRL2014} on the notion of (i) incoherent states and (ii) incoherent operations. In \cite{PlenioPRL2014} the authors consider a $d$-dimensional Hilbert space spanned by the basis states $\ket{i}$, with $i=1,2,...d$, and define as incoherent states all density matrices of the form $\sum_{i=1}^{d}\delta_i\ket{i}\bra{i}$, where the non-negative weights $\delta_i$ sum up to unity. 

In our case we consider a $d$-dimensional Hilbert space of the radical-pair under consideration, which is spanned by $d$ states, but we do not care about a "global" coherence, but only about coherence between the singlet and triplet subspaces, hence the following definition.
\subsection{Definition of S-T incoherent states}
Singlet-triplet incoherent radical-pairs are those for which the radical-pair density matrix $\rho$ has the property that $\rho=\qs\rho\qs+\qt\rho\qt\equiv\rho_{\rm SS}+\rho_{\rm TT}$, or equivalently, those for which the density matrix has the property that $\qs\rho\qt+\qt\rho\qs\equiv\rho_{\rm ST}+\rho_{\rm TS}=0$.
This definition straightforwardly results from the identity \eqref{ident}. Let all incoherent states (for a particular radical-pair Hilbert space) define the set ${\cal I}$.
\subsection{Definition of S-T incoherent operations}
A set of Kraus operators $K_n$, with $\sum_n K_n^\dagger K_n=\mathbbmtt{1}$, are called incoherent \cite{PlenioPRL2014} if for all $n$ it is $K_n{\cal I}K_n^\dagger\subset{\cal I}$. In our case, the two projectors $\qs$ and $\qt$ qualify as a set of incoherent operations, since $\qs\qs+\qt\qt=\qs+\qt=\mathbbmtt{1}$, and for any $\rho\subset{\cal I}$, i.e. $\rho=\rho_{\rm SS}+\rho_{\rm TT}$ it is $\qs\rho\qs=\rho_{\rm SS}\subset{\cal I}$ and $\qt\rho\qt=\rho_{\rm TT}\subset{\cal I}$. The pair $\qs$ and $\qt$ are incoherent operators of central significance for discussing S-T coherence, but by no means are they the only ones.

For example, any operators acting only on the nuclear spins are S-T incoherent operators. Consider $K_n=\mathbbmtt{1}\otimes k_n$, where now $\mathbbmtt{1}$ is the unit matrix in the electronic spin subspace (4-dimensional), and $k_n$ are trace-preserving Kraus operators acting in the nuclear spin subspace and satisfying $\sum_n k_n^\dagger k_n=\mathbbmtt{1}$. Take a density matrix $\rho\subset{\cal I}$. Then $\rho=\qs\rho\qs+\qt\rho\qt$, so that $K_n\rho K_n^\dagger=K_n\qs\rho\qs K_n^\dagger+K_n\qt\rho\qt K_n^\dagger$. It can be readily seen that $\qs$ and $\qt$ commute with $K_n$, therefore $K_n\qs\rho\qs K_n^\dagger+K_n\qt\rho\qt K_n^\dagger=\qs K_n\rho K_n^\dagger\qs+\qt K_n\rho K_n^\dagger\qt$. Since $K_n$ define a trace-preserving map, $K_n\rho K_n^\dagger$ is also a physical (yet unnormalized) density matrix, call it $R$, hence finally $K_n\qs\rho\qs K_n^\dagger+K_n\qt\rho\qt K_n^\dagger=\qs R\qs+\qt R\qt\subset{\cal I}$. That is, we have shown that for any $\rho\subset{\cal I}$ it will be $K_n\rho K_n^\dagger\subset{\cal I}$, thus $K_n$ are incoherent operators.
\subsection{Definition of S-T coherence quantifier based on relative entropy}
For any radical-pair density matrix $\rho$ we define the singlet-triplet coherence quantifier as
\beq
{\cal C}\llbracket\rho\rrbracket={\cal S}\llbracket\qs\rho\qs+\qt\rho\qt\rrbracket-{\cal S}\llbracket\rho\rrbracket,\label{Cdef}
\eeq
where ${\cal S}\llbracket r\rrbracket$ is the von-Neuman entropy of the density matrix $r$. Since radical-pair reactions are non-trace preserving, the trace of the radical-pair state $\rho$ is in general $0\leq\tr\{\rho\}\leq 1$. For the definition \eqref{Cdef} to work, we first need to normalize the radical-pair state $\rho$ with $\tr\{\rho\}$ (see Appendix of \cite{PRE2017} for a relevant discussion). In the following we will always imply that whenever we calculate ${\cal C}\llbracket\rho\rrbracket$ we do so for radical-pair density matrices that have been appropriately normalized to have unit trace.

At first sight, the quantum relative entropy is not present in the definition \eqref{Cdef}. However, equation \eqref{Cdef} readily follows by first defining \cite{PlenioPRL2014}
\beq
{\cal C}\llbracket\rho\rrbracket=\min_{\delta\subset{\cal I}}S\llbracket\rho ||\delta\rrbracket,
\eeq
where now $S\llbracket\rho ||\delta\rrbracket\equiv \tr\{\rho\log\rho\}-\tr\{\rho\log\delta\}$ is the quantum relative entropy of the radical-pair density matrices $\rho$ and $\delta$. Indeed, by denoting 
\beq
\hat{\rho}=\qs\rho\qs+\qt\rho\qt,
\eeq
and for $\delta\subset{\cal I}$ it is \cite{PlenioPRL2014} $S\llbracket\rho ||\delta\rrbracket=S\llbracket\hat{\rho}||\delta\rrbracket+S\llbracket\hat{\rho}\rrbracket-S\llbracket\rho\rrbracket$. However, since the quantum relative entropy is always positive or zero, $S\llbracket\hat{\rho}||\delta\rrbracket\geq 0$, it is seen that $\min_{\delta\subset{\cal I}}S\llbracket\hat{\rho}||\delta\rrbracket=0$, the minimum obviously taking place for $\delta=\hat{\rho}$, since it is known that for any density matrix $r$ it is $S\llbracket r||r\rrbracket=0$. Hence ${\cal C}\llbracket\rho\rrbracket=S\llbracket\rho ||\hat{\rho}\rrbracket=S\llbracket\hat{\rho}\rrbracket-S\llbracket\rho\rrbracket$.
\subsection{Properties of ${\cal C}\llbracket\rho\rrbracket$}
We here present the basic properties of definition \eqref{Cdef}.
\subsubsection{Value of ${\cal C}\llbracket\rho\rrbracket$ for incoherent states}
\noindent As required from any acceptable coherence quantifier, for all $\rho\subset{\cal I}$ it is ${\cal C}\llbracket\rho\rrbracket=0$.\newline\newline
\textbf{\emph{Proof}} This follows trivially from the definition \eqref{Cdef}, since if $r\subset{\cal I}$ it is $r=\qs r\qs+\qt r\qt$, hence ${\cal C}\llbracket r\rrbracket={\cal S}\llbracket r\rrbracket-{\cal S}\llbracket r\rrbracket=0$. 
\subsubsection{Minimum value of ${\cal C}\llbracket\rho\rrbracket$}
\noindent For any $\rho$ it is ${\cal C}\llbracket\rho\rrbracket\geq 0$.\newline\newline
\textbf{\emph{Proof}} The proof follows from the fact that under a nonselective (or blind) measurement the entropy does not decrease. This can be found in pp. 75 of \cite{Jacobs} and pp.92 of \cite{Breuer}.
In our case, defining a measurement by the Kraus operators $\qs$ and $\qt$, which satisfy $\qs\qs^\dagger+\qt\qt^\dagger=\qs\qs+\qt\qt=\qs+\qt=\mathbbmtt{1}$, the non-selective post-measurement state is given by $\qs\rho\qs+\qt\rho\qt$. Indeed, the measurement results are $q_s=1$ or $q_s=0$, taking place with probabilities $p_{\rm S}=\tr\{\rho\qs\}$ and $p_{\rm T}=\tr\{\rho\qt\}$, with the selective post-measurement states being $\rho_{\rm S}=\qs\rho\qs/p_{\rm S}$ and $\rho_{\rm T}=\qt\rho\qt/p_{\rm T}$, respectively. The non-selective post-measurement state is $p_{\rm S}\rho_{\rm S}+p_{\rm T}\rho_{\rm T}$ which indeed equals $\qs\rho\qs+\qt\rho\qt$. Since, by the theorem found in \cite{Jacobs,Breuer} it is ${\cal S}\llbracket\qs\rho\qs+\qt\rho\qt\rrbracket\geq {\cal S}\llbracket\rho\rrbracket$, it follows that indeed ${\cal C}\llbracket\rho\rrbracket\geq 0$.
\subsubsection{Maximum value of ${\cal C}\llbracket\rho\rrbracket$}
\noindent For any $\rho$ it is ${\cal C}\llbracket\rho\rrbracket\leq 1$.\newline\newline
\textbf{\emph{Proof}} First, the quantum relative entropy is jointly convex in both of its arguments \cite{Ohya}, i.e. 
\begin{align}
&S\llbracket\lambda\rho_1+(1-\lambda)\rho_2||\lambda\sigma_1+(1-\lambda)\sigma_2\rrbracket\leq\nonumber\\
&\lambda S\llbracket\rho_1||\sigma_1\rrbracket+(1-\lambda)S\llbracket\rho_2||\sigma_2\rrbracket
\end{align}
for any $\lambda\in [0,1]$. Applying this property for $\rho_1=\rho_2=\rho$, $\sigma_1=\rho_{\rm S}$, $\sigma_2=\rho_{\rm T}$, $\lambda=p_{\rm S}$, and given that $\hat{\rho}=p_{\rm S}\rho_{\rm S}+p_{\rm T}\rho_{\rm T}$, it follows that  
\begin{align}
S\llbracket\rho ||\hat{\rho}\rrbracket&=S\llbracket\rho ||p_{\rm S}\rho_{\rm S}+p_{\rm T}\rho_{\rm T}\rrbracket\nonumber\\
&\leq p_{\rm S}S\llbracket\rho ||\rho_{\rm S}\rrbracket+p_{\rm T}S\llbracket\rho ||\rho_{\rm T}\rrbracket\label{convex}
\end{align}

Now, the interpretation of the quantum relative entropy $S\llbracket\rho ||\sigma\rrbracket$ is \cite{Vedral} the extent to which one can distinguish two different states $\rho$ and $\sigma$, in particular by a series of quantum measurements and their resulting statistics. Let us first consider $S\llbracket\rho ||\rho_{\rm S}\rrbracket$. This reflects the extent to which by doing some measurement on $\rho$ we can use the measurement statistics to distinguish $\rho$ from $\rho_{\rm S}$. We can choose as measurement the measurement of $\qs$, the result of which can be either 0 or 1. Clearly if the state we were measuring was $\rho_{\rm S}=\qs\rho\qs/\tr\{\rho\qs\}$, we would obtain 1 for every measurement performed in $N$ identically prepared systems. But if the state of each of those identical copies is $\rho$, measuring $\qs$ we will obtain 1 only some of the times. The probability to obtain $1$ in all $N$ such measurements will clearly be $p_{\rm S}^N$. This is the probability that $\rho_{\rm S}$ would "pass" our test and we would confuse the actual state $\rho$ with $\rho_{\rm S}$. But it is known \cite{Vedral} that for an optimal measurement using $N$ identical copies of our system, the probability that we will mistakenly confuse $\rho$ for $\rho_{\rm S}$ is $2^{-NS\llbracket\rho ||\rho_{\rm S}\rrbracket}$. Our choice of measurement is not necessarily optimal, hence $2^{-NS\llbracket\rho ||\rho_{\rm S}\rrbracket}\leq p_{\rm S}^N$, or $S\llbracket\rho ||\rho_{\rm S}\rrbracket\leq -\log [p_{\rm S}]$. We can similarly show that $S\llbracket\rho ||\rho_{\rm T}\rrbracket\leq -\log [p_{\rm T}]=-\log[1-p_{\rm S}]$. Finally, using the inequality \eqref{convex} we get
\beq
{\cal C}\llbracket\rho\rrbracket\leq\mathbb{H}[p_{\rm S},p_{\rm T}],\label{bound}
\eeq
where $\mathbb{H}[p_{\rm S},p_{\rm T}]=-p_{\rm S}\log p_{\rm S}-p_{\rm T}\log p_{\rm T}=-p_{\rm S}\log p_{\rm S}-(1-p_{\rm S})\log(1-p_{\rm S})$ is the Shannon entropy of the pair of probabilities $\{p_{\rm S}, p_{\rm T}\}$. Since this Shannon entropy has maximum 1 (when the logarithms are calculated with base 2), the maximum occuring for $p_{\rm S}=p_{\rm T}=1/2$, we finally show that the maximum of ${\cal C}\llbracket\rho\rrbracket$ is 1.
\subsubsection{Singlet-triplet coherence of pure states}
\noindent Pure radical-pair states $\ket{\psi}$ saturate the bound \eqref{bound}, i.e. it is ${\cal C}\llbracket\ket{\psi}\bra{\psi}\rrbracket=\mathbb{H}[p_{\rm S},p_{\rm T}]$, where $p_{\rm S}=\tr\{\ket{\psi}\bra{\psi}\qs\}$.\newline\newline
\textbf{\emph{Proof}}
The most general pure state of a radical-pair can be written as 
\beq
\ket{\psi}=\alpha_{\rm s}\ket{\rm s}\otimes\ket{\chi_{\rm s}}+\sum_{j=-1}^{1}\beta_{j}\ket{{\rm t}_j}\otimes\ket{\chi_j},\label{pure} 
\eeq
where $\ket{\chi_{\rm s}}$ and $\ket{\chi_j}$ are normalized nuclear spin states "living" in a nuclear spin space of dimension $d_{\rm nuc}$ (see Sec. II A). 
Setting $\rho=\ket{\psi}\bra{\psi}$, in order to calculate ${\cal C}\llbracket\rho\rrbracket$, we need to calculate the entropies ${\cal S}\llbracket\rho\rrbracket$ and ${\cal S}\llbracket\hat{\rho}\rrbracket$. The former is zero since $\rho$ is a pure state.
To calculate the latter, we write
\begin{align}
\hat{\rho}&=\qs\rho\qs+\qt\rho\qt\nonumber\\
&=|\alpha_{\rm s}|^2\ket{\rm s}\bra{\rm s}\otimes\ket{\chi_{\rm s}}\bra{\chi_{\rm s}}+\ket{\phi_{\rm T}}\bra{\phi_{\rm T}}
\end{align}
where for brevity we set $\ket{\phi_{\rm T}}=\sum_{j=-1}^{1}\beta_{j}\ket{{\rm t}_j}\otimes\ket{\chi_j}$ for the triplet-subspace component of the most general pure state $\ket{\psi}$. The matrix $\hat{\rho}$ clearly has a block-diagonal form, one block being the singlet and the other the triplet subspace. To calculate ${\cal S}\llbracket\hat{\rho}\rrbracket$ we need to find the eigenvalues of $\hat{\rho}$. They are easily obtained by finding the eigenvectors and corresponding eigenvalues of $\hat{\rho}$ by construction. 

For example, the state $\ket{\rm s}\otimes\ket{\chi_{\rm s}}$ is an eigenvector of the singlet block diagonal of $\hat{\rho}$ with eigenvalue $|\alpha_{\rm s}|^2=\tr\{\rho\qs\}=p_{\rm S}$. Remaining in this singlet subspace block-diagonal, we can span the nuclear spin space with $d_{\rm nuc}$ orthogonal basis states, one being $\ket{\chi_{\rm s}}$ itself. Hence the other $d_{\rm nuc}-1$ eigenvalues of the singlet block-diagonal of $\hat{\rho}$ are zero. Similarly, the unnormalized state $\ket{\phi_{\rm T}}$ is an eigenstate of $\ket{\phi_{\rm T}}\bra{\phi_{\rm T}}$ with eigenvalue $\braket{\phi_{\rm T}|\phi_{\rm T}}=|\beta_{-1}|^2+|\beta_{0}|^2+|\beta_{1}|^2=\tr\{\rho\qt\}=p_{\rm T}$. We can clearly span this triplet subspace with $3d_{\rm nuc}$ orthogonal basis states, one of which is $\ket{\phi_{\rm T}}$ itself. Hence the other eigenvalues of $\ket{\phi_{\rm T}}\bra{\phi_{\rm T}}$ are zero. Thus, the state $\hat{\rho}$ has two nonzero eigenvalues, $p_{\rm S}=\tr\{\rho\qs\}$ and $p_{\rm T}=\tr\{\rho\qt\}=1-p_{\rm S}$, hence the coherence measure of the most general pure radical-pair state \eqref{pure} is exactly equal to $\mathbb{H}[p_{\rm S},p_{\rm T}]$.
\subsubsection{States with maximum singlet-triplet coherence}
It readily follows that the states having maximum S-T coherence, equal to 1, are all pure states of the form $\ket{\psi}_{\rm max C}={1\over\sqrt{2}}\ket{\rm s}\otimes\ket{\chi_{\rm s}}+\sum_{j=-1}^1c_j\ket{\rm t_j}\otimes\ket{\chi_j}$, where $\sum_{j=-1}^1|c_j|^2=1/2$, while $\ket{\chi_{\rm s}}$ and $\ket{\chi_j}$ are arbitrary normalized nuclear spin states.
\subsubsection{Connection with quantum uncertainty}
For the general pure state $\ket{\psi}$ of Eq. \eqref{pure} it can be easily seen that the quantum uncertainty of $\qs$, given by $\Delta q_{\rm s}\equiv\sqrt{\bra{\psi}\qs^2\ket{\psi}-\bra{\psi}\qs\ket{\psi}^2}=\sqrt{\bra{\psi}\qs\ket{\psi}-\bra{\psi}\qs\ket{\psi}^2}$, is
$\Delta q_{\rm s}=\sqrt{|\alpha_{\rm s}|^2(1-|\alpha_{\rm s}|^2)}$. Evidently, $\Delta q_{\rm s}$ is maximized for $|\alpha_{\rm s}|=1/\sqrt{2}$, i.e. the maximally coherent pure states $\ket{\psi}_{\rm max C}$ also have maximum uncertainty in their singlet (or equivalently triplet) character. This is intuitively satisfactory, since thinking at the level of a simple qubit, we intuitively relate the maximum coherence state $(\ket{0}+\ket{1})/\sqrt{2}$ with the fact that this state is maximally uncertain regarding a measurement in the computational basis $\{\ket{0},\ket{1}\}$.
\subsubsection{Additional comments	}
All other conditions for an acceptable measure of coherence, like (i) monotonicity under incoherent completely positive and trace preserving maps, (ii) monotonicity under selective measurements on average and (iii) convexity, are automatically satisfied as has been shown for the relative entropy measure defined in \cite{PlenioPRL2014}.
\subsection{Examples}
\subsubsection{Fictitious radical-pair with no nuclear spins}
We will first consider a simple and analytic example of a fictitious radical-pair with no nuclear spins, hence a four-dimensional spin space spanned by $\ket{\rm s}$ and $\ket{\rm t_j}$, with $j=0,\pm 1$. In fact, high magnetic fields, at which the Zeeman terms dominate hyperfine interactions, readily lead to this approximation and in such cases S-T mixing is driven by a difference in the g-factor for the electronic spins in the two radicals \cite{Molin}. Explicitly, consider a Hamiltonian ${\cal H}=\omega_1 s_{1z}+\omega_{2} s_{2z}$, where $\omega_1\neq\omega_2$ are the Larmor frequencies of the electrons in the two radicals. If the initial state is $\ket{\psi_0}=\ket{\rm s}$, it is easily seen that $\ket{\psi_t}=\cos{{\Omega t}\over 2}\ket{\rm s}-i\sin{{\Omega t}\over 2}\ket{\rm t_0}$, where $\Omega=\omega_1-\omega_2$. Thus the singlet and triplet probabilities are $p_{\rm S}=\cos^2{{\Omega t}\over 2}$ and $p_{\rm T}=\sin^2{{\Omega t}\over 2}$, respectively. 

\begin{figure}[t]
\begin{center}
\includegraphics[width=8 cm]{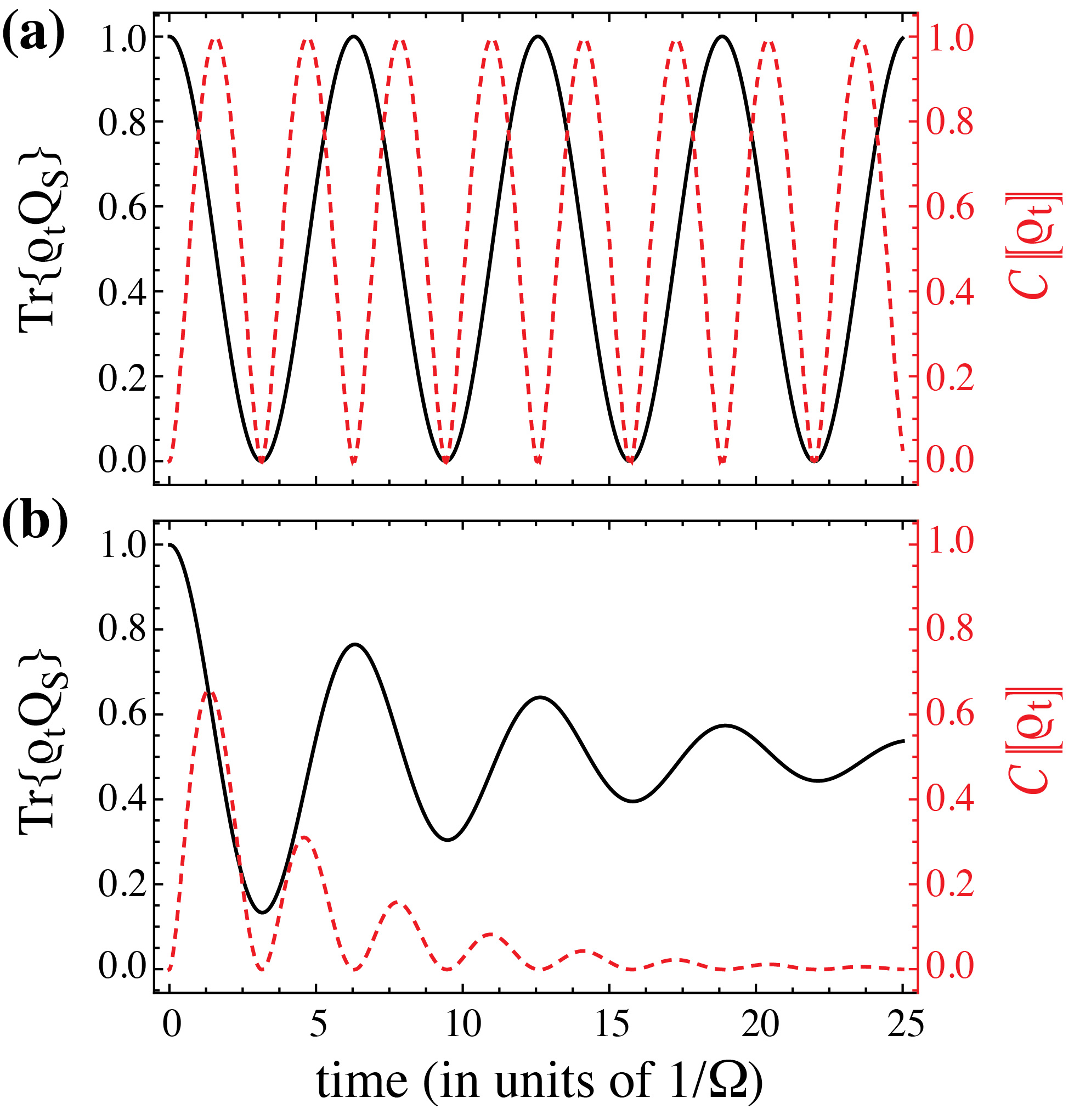}
\caption{Singlet probability $\tr\{\rho_t\qs\}$ (black solid line), and coherence measure ${\cal C}\llbracket\rho_t\rrbracket$ (red dashed line), for a two-electron system evolving by ${\cal H}=\omega_1 s_{1z}+\omega_2 s_{2z}$, with initial state at $t=0$ being $\ket{\rm s}$. The S-T Rabi frequency is $\Omega=\omega_1-\omega_2$. (a) Unitary evolution driven only by ${\cal H}$. S-T coherence is zero whenever the state is a pure singlet ($\tr\{\rho_t\qs\}=1$) or a pure triplet ($\tr\{\rho_t\qs\}=0$), and maximum (equal to 1) when the state is $(\ket{\rm s}-i\ket{\rm t_0})/\sqrt{2}$. (b) Non-unitary evolution caused by S-T dephasing at a rate $K_d=\Omega/5$.}
\label{ex_coh}
\end{center}
\end{figure}

The state $\rho_t=\ket{\psi_t}\bra{\psi_t}$ is pure, hence $S\llbracket\rho_t\rrbracket=0$. The incoherent state $\hat{\rho}_t=\qs\rho\qs+\qt\rho\qt$ is $\hat{\rho}_t=p_{\rm S}\ket{\rm s}\bra{\rm s}+p_{\rm T}\ket{\rm t_0}\bra{\rm t_0}$. It readily follows that the eigenvalues of $\hat{\rho}_t$ are $p_{\rm S}$ and $p_{\rm T}$, hence $S\llbracket\hat{\rho}_t\rrbracket=-p_{\rm S}\log p_{\rm S}-p_{\rm T}\log p_{\rm T}$. Thus the coherence measure for $\rho_t$ is ${\cal C}\llbracket\rho_t\rrbracket=S\llbracket\hat{\rho}_t\rrbracket-S\llbracket\rho_t\rrbracket=-p_{\rm S}\log p_{\rm S}-p_{\rm T}\log p_{\rm T}$.
In Fig. \ref{ex_coh}a we plot the time evolution of the singlet probability $p_{\rm S}=\tr\{\rho_t\qs\}$ and ${\cal C}\llbracket\rho_t\rrbracket$. Evidently, ${\cal C}\llbracket\rho_t\rrbracket$ is zero when $\ket{\psi_t}$ is a pure singlet or a pure triplet, and reaches its maximum value of 1 in between the maxima of $p_{\rm S}$, i.e. at those times where we have the most uncertain coherent superposition of $\ket{\rm s}$ and $\ket{\rm t_0}$, of the form ${1\over\sqrt{2}}(\ket{\rm s}-i\ket{\rm t_0})$.

We can now introduce an S-T dephasing through the operation $\rho\rightarrow\rho-K_d dt(\qs\rho\qt+\qt\rho\qs)$, i.e. by removing from $\rho$ its coherent part $\rho_{\rm ST}+\rho_{\rm TS}$ at a rate $K_d$ (we will elaborate more about this in the following section). In the presence of such an S-T dephasing mechanism, a pure initial state necessarily evolves into a mixed state, which in general satisfies the master equation $d\rho/dt=-i[{\cal H},\rho]-K_d(\qs\rho\qt+\qt\rho\qs)$. This is easy to solve analytically in the considered example, since the problem is essentially reduced to a two dimensional system spanned by $\ket{s}$ and $\ket{\rm t_0}$. For $K_d<2|\Omega|$ the off-diagonal density matrix elements are oscillatory and decay exponentially at a rate $K_d/2$. An analytic expression can be obtained for $\rho_t$, but it is a bit cumbersome. For $K_d\ll 2|\Omega|$ an excellent approximation is $\rho_t={1\over 2}(1+e^{-K_d t/2}\cos\Omega t)\ket{\rm s}\bra{\rm s}+{1\over 2}(1-e^{-K_d t/2}\cos\Omega t)\ket{\rm t_0}\bra{\rm t_0}+{i\over 2}e^{-K_d t/2}\sin\Omega t\ket{\rm s}\bra{\rm t_0}-{i\over 2}e^{-K_d t/2}\sin\Omega t\ket{\rm t_0}\bra{\rm s}$. 

The singlet probability is now $\tr\{\rho_t\qs\}={1\over 2}(1+e^{-K_d t/2}\cos\Omega t)$. The eigenvalues of $\rho_t$ and $\hat{\rho}_t$ are $e_1={1\over 2}(1-e^{-K_d t/2})$, $e_2={1\over 2}(1+e^{-K_d t/2})$ and $\hat{e}_1={1\over 2}(1-e^{-K_d t/2}\cos\Omega t)$, $\hat{e}_2={1\over 2}(1+e^{-K_d t/2}\cos\Omega t)$, respectively. Thus we can readily calculate the entropies $S\llbracket\rho_t\rrbracket$ and $S\llbracket\hat{\rho}_t\rrbracket$, and from them ${\cal C}\llbracket\rho_t\rrbracket=S\llbracket\hat{\rho}_t\rrbracket-S\llbracket\rho_t\rrbracket=-\hat{e}_1\log\hat{e}_1-\hat{e}_2\log\hat{e}_2+e_1\log e_1+e_2\log e_2$. These analytic results are shown in Fig. \ref{ex_coh}b.
\subsubsection{Incoherent operations on nuclear spins}
The phenomenological wealth of coherence phenomena increases dramatically by considering realistic radical-pairs involving one ore more nuclear spins. In particular, plots similar to Fig. \ref{ex_coh} can also be produced by introducing just one nuclear spin and calculating the evolution of S-T coherence in various scenarios. However, here we will explicitly mention as a second example a more subtle effect having to do with the fact that the pair $\{\qs,\qt\}$ are not the only incoherent operations, as mentioned in Sec. III B. In the presence of nuclear spins, there is a mechanism irrelevant to the electronic spins, by which S-T dephasing can be produced. {\it This can happen when there is electronic-nuclear entanglement}. 

Consider in particular a pure radical-pair state of the form $\ket{\psi}=(\ket{\rm s}\otimes\ket{\Uparrow}+\ket{\rm t_0}\otimes\ket{\Downarrow})/\sqrt{2}$. Ignoring for the moment the electronic states $\ket{\rm t_1}$ and  $\ket{\rm t_{-1}}$, so that the electron spin subspace "looks" two-dimensional, the state $\ket{\psi}$ is clearly a maximally entangled state of two qubits, one spanned by the states $\ket{\rm s}$ and $\ket{\rm t_0}$ and the other by $\ket{\Uparrow}$ and $\ket{\Downarrow}$. Now, suppose we perform a measurement on the nuclear spin only, e.g. in the $z$-basis, using the Kraus operators $K_1=\mathbbmtt{1}\otimes\ket{\Uparrow}\bra{\Uparrow}$ and $K_2=\mathbbmtt{1}\otimes\ket{\Downarrow}\bra{\Downarrow}$, which evidently leave the electrons untouched. Yet, because of the electron-nucleus entanglement, these S-T incoherent operators will also damp S-T coherence, as easily seen by setting $\rho=\ket{\psi}\bra{\psi}$ and considering $K_1\rho K_1+K_2\rho K_2={1\over 2}\ket{\rm s}\bra{\rm s}\otimes\ket{\Uparrow}\bra{\Uparrow}+{1\over 2}\ket{\rm t_0}\bra{\rm t_0}\otimes\ket{\Downarrow}\bra{\Downarrow}$. The state $\rho$ is maximally S-T coherent, while $K_1\rho K_1+K_2\rho K_2$ has zero S-T coherence.
\section{Singlet-triplet coherence as a resource for magnetoreception}
We will now use the formally introduced quantifier to show that S-T coherence is a resource for biological magnetoreception. This is not an innocuous statement, and care should be exercised in arguing for S-T coherence providing an operational advantage to the chemical compass. 

It is known that an anisotropic hyperfine coupling can render the radical-pair reaction yields dependent on the angle $\phi$ of the external magnetic field with respect to a molecule-fixed coordinate frame. As is usually the case, we consider a single-nuclear-spin radical pair, with Hamiltonian 
\beq
{\cal H}=\mathbf{s}_1\cdot\mathbf{A}\cdot\mathbf{I}+\omega\cos\phi(s_{1x}+s_{2x})+\omega\sin\phi(s_{1y}+s_{2y}),\label{Ham}
\eeq
where $\mathbf{A}$ is the hyperfine tensor coupling the electron spin $\mathbf{s}_1$ of one radical with the single nuclear spin $\mathbf{I}$ of that radical, and $\omega$ the external magnetic field lying on the $x-y$ plane and producing the Zeeman terms of the two electronic spins (we omit the nuclear Zeeman term). We will now calculate the singlet reaction yield as a function of the angle $\phi$, always starting at $t=0$ with a singlet state for the electrons and a fully mixed nuclear spin state, $\rho_0=\qs/\tr\{\qs\}$. As mentioned in Sec. II B, we assume that $\ks=\kt=k$, in which case the quantum dynamics of the radical-pair reaction are simplified, and the differences between our master equation and Haberkorn's are less exacerbated. In particular, when $\ks=\kt=k$ it is $d\rho/dt=e^{-kt}R$, with $dR/dt=-i[{\cal H},R]-k(\qs R\qt+\qt R\qs)$ according to our theory, whereas $dR/dt=-i[{\cal H},R]$ according to Haberkorn's, i.e. we have an additional dephasing term $-k(\qs R\qt+\qt R\qs)$ inherent in our description of the dynamics. In the following we will anyhow use a significantly stronger dephasing term of the form $-K_d(\qs R\qt+\qt R\qs)$, with $K_d\gg k$, hence it really does not matter for this discussion which of the two master equations we use \cite{note}.

\begin{figure}[t]
\begin{center}
\includegraphics[width=7.5 cm]{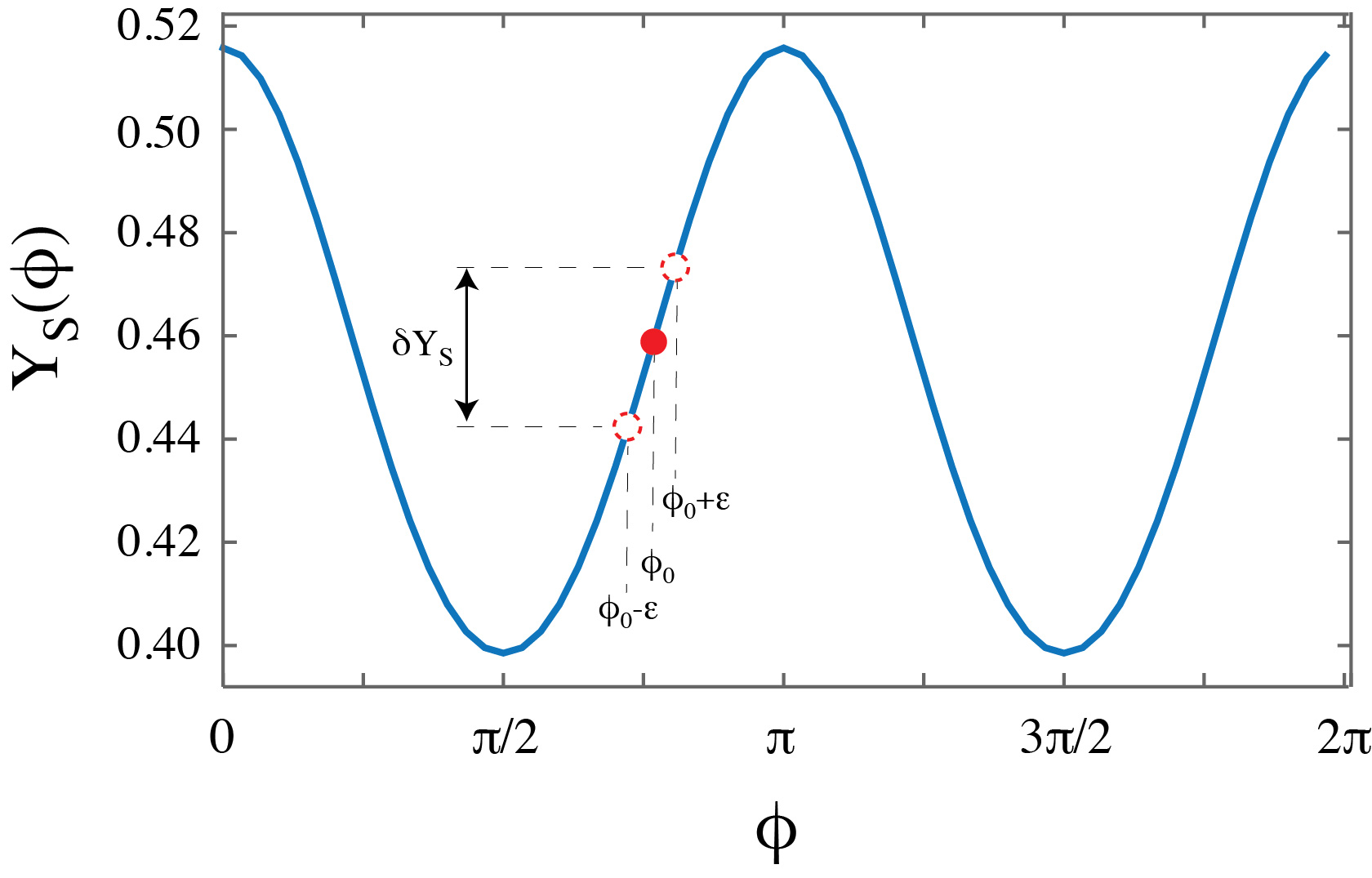}
\caption{Example of the angular dependence of the singlet reaction yield, for the Hamiltonian of Eq. \eqref{Ham}, having a diagonal hyperfine tensor with $A_{yy}=A_{zz}=0$, $A_{xx}/k=10$ and $\omega/k=1$. The figure of merit of the compass is the value $\delta Y_{\rm S}$, which quantifies the highest slope of $Y_{\rm S}$ versus $\phi$.}
\label{fig3}
\end{center}
\end{figure}

We now calculate the singlet reaction yield, $Y_{\rm S}(\phi)=\int_{0}^{\infty}dt k\tr\{\rho_t\qs\}$, and plot it as a function of $\phi$ in order to define the figure of merit for the magnetic compass. We remind the reader that the yield $Y_{\rm S}$ is a function of $\phi$ because the master equation evolving $\rho_0$ into $\rho_t$  depends on $\phi$ through the Hamiltonian ${\cal H}$. An example of such a $\phi$-dependence of $Y_{\rm S}$ is shown in Fig. \ref{fig3}. As is usually the case, we define the figure of merit by
\beq
\delta Y_{\rm S}=\max_{\phi_0}|Y_{\rm S}(\phi_0+\epsilon)-Y_{\rm S}(\phi_0-\epsilon)|\label{dys}
\eeq
\subsection{Correlation of $\delta Y_{\rm S}$ and $\overline{{\cal C}}$}
We will now explore the connection between S-T coherence quantified by ${\cal C}$ and the figure of merit $\delta Y_{\rm S}$. Since ${\cal C}$ depends on the time-dependent density matrix $\rho_t$, we define a mean value of ${\cal C}\llbracket\rho_t\rrbracket$ along the whole reaction as $\overline{\cal C}=\int_{0}^{\infty}dt ke^{-kt}{\cal C}\llbracket\rho_t\rrbracket$. We remind the reader that in order to calculate ${\cal C}\llbracket\rho_t\rrbracket$ using the definition \eqref{Cdef}, we always have to first normalize $\rho_t$ by $\tr\{\rho_t\}$ (since this trace changes with time due to the reaction), and then calculate the entropies in \eqref{Cdef}. We use the Hamiltonian of Eq. \eqref{Ham}, and for completeness we add an exchange term of the form $-J\mathbf{s}_1\cdot\mathbf{s}_2$. We use a diagonal hyperfine tensor, randomizing all three diagonal elements $A_{jj}$, with $j=x,y,x$. We also randomize the exchange coupling $J$. For each set of parameters $A_{xx}$, $A_{yy}$, $A_{zz}$ and $J$ we calculate $\delta Y_{\rm S}$ and $\overline{\cal C}$. 
\begin{figure*}[ht]
\begin{center}
\includegraphics[width=16.5 cm]{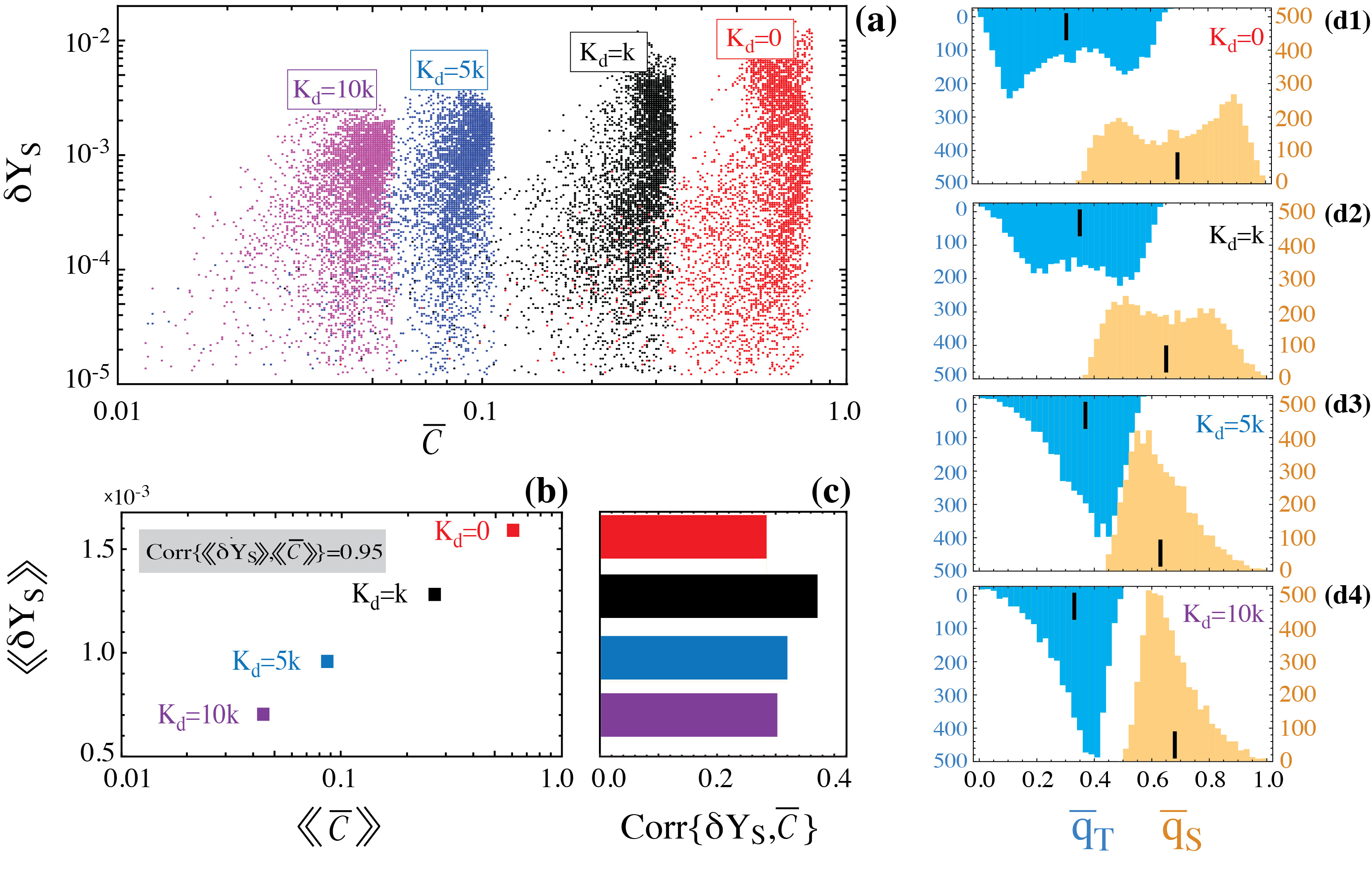}
\caption{(a) Correlation map of the compass' figure of merit $\delta Y_{\rm S}$ versus the S-T coherence averaged along the reaction, $\overline{\cal C}$, as calculated from the master equation $\rho=e^{-kt}R$, with $dR/dt=-i[{\cal H},R]-K_{d}(R_{\rm ST}+R_{\rm TS})$, where ${\cal H}$ is the Hamiltonian of Eq. \eqref{Ham} including an exchange term. Each point results from a random set of parameters $0\leq A_{xx},A_{yy},A_{zz}\leq 10k$,  $-10\leq J/k\leq 10$ and $\omega/k=1$. The dephasing rate $K_d$ was given the values $K_d=0$, $K_d=k$, $K_d=5k$ and $K_d=10k$, with 5000 points for each. Initial state was always a singlet with mixed nuclear spin, $\rho_0=\qs/\tr\{\qs\}$. (b) Mean value of $\delta Y_{\rm S}$ versus mean value of $\overline{\cal C}$ over all 5000 points for each value of $K_d$. (c) Correlation coeffecient between $\delta Y_{\rm S}$ and $\overline{\cal C}$ for each value of $K_d$. (d1)-(d4) Distribution of $Y_{\rm S}$ and $Y_{\rm T}$ for each value of $K_d$. Also shown is the mean of each distribution (black lines). The distributions are seen to become more narrow at high $K_d$. This is because the dephasing operation is equivalent to a quantum measurement with Kraus operators $\qs$ and $\qt$, the narrowing reflecting the measurement-induced localization of the radical-pair's state.}
\label{corr}
\end{center}
\end{figure*}
Additionally, we calculate the mean values along the reaction of the singlet and triplet expectation values, $\langle\qs\rangle_t=\tr\{\rho_t \qs\}$ and $\langle\qt\rangle_t=\tr\{\rho_t \qt\}$, which are nothing other than the singlet and triplet reaction yields, $Y_{\rm S}=\int_{0}^{\infty}dt k\tr\{\rho_t \qs\}$ and  $Y_{\rm T}=\int_{0}^{\infty}dt k\tr\{\rho_t \qt\}$, respectively. The mean values $\overline{\cal C}$, $Y_{\rm S}$ and $Y_{\rm T}$ are calculated for the particular $\phi_0$ maximizing $\delta Y_{\rm S}$ (see Eq. \eqref{dys}).

Now, the density matrix evolution is $\rho_t=e^{-kt}R$, where the density matrix $R$ satisfies the master equation $dR/dt=-i[{\cal H},R]-K_d(\qs R\qt+\qt R\qs)$. We repeat the aforementioned calculations for four different values of the dephasing rate $K_d$, in particular for $K_d=0$, $K_d=k$, $K_d=5k$ and $K_d=10k$. That is, we vary $K_d$ appreciably in order to explore the effect of suppressing S-T coherence. The main results of this simulation, making a clear case that S-T coherence is indeed a resource for magnetoreception, are shown in Fig. \ref{corr}. We will make a number of observations, and then provide their interpretation in the following subsection.\newline
\noindent
{\bf (1)} In Fig. \ref{corr}a we show for each value of the dephasing rate $K_d$ the distribution of 5000 pairs of $\delta Y_{\rm S}$ and $\bar{\cal C}$. We first note that by increasing $K_d$, the distribution moves to smaller $\delta Y_{\rm S}$ and smaller $\bar{\cal C}$. This is more evident in Fig. \ref{corr}b, where we plot the mean value of these two quantities over the sample of 5000 points.\newline
\noindent
{\bf (2)} Irrespective of $K_d$, there seems to be an appreciable correlation coefficient (around 0.3) between $\delta Y_{\rm S}$ and $\overline{\cal C}$, as shown in Fig. \ref{corr}c. Moreover, the correlation between the sample means $\langle\langle\delta Y_{\rm S}\rangle\rangle$ and $\langle\langle\bar{\cal C}\rangle\rangle$, presented in Fig. \ref{corr}b, is much larger, and has the value 0.95.\newline
\noindent
{\bf (3)} Finally, in Figs. \ref{corr}(d1)-\ref{corr}(d4) we plot the distribution of the singlet and triplet character of the radical-pair state along the reaction, quantified by the singlet and triplet reaction yields $Y_{\rm S}$ and $Y_{\rm T}$, respectively.
\subsection{Interpretation}
We will now interpret the aforementioned observations. The first question to ask is, {\it does singlet-triplet coherence provide a quantum advantage to the operation of the compass}? The answer should clearly be affirmative, because of three facts: (a) Due to the correlation between $\delta Y_{\rm S}$ and $\bar{\cal C}$ at a {\it specific} value of $K_d$, large values of S-T coherence $\overline{\cal C}$ are on average connected with large figures of merit $\delta Y_{\rm S}$ for the compass.  (b) Strong S-T dephasing produced by increasing $K_d$ leads to small values of $\bar{\cal C}$ {\it and} small values of $\delta Y_{\rm S}$. (c) {\it At the same time} the average singlet and triplet populations, as seen in Figs. \ref{corr}(d1)-\ref{corr}(d4), are not affected by increasing $K_d$. In other words, we have a process by which an initially singlet radical-pair state is coherently transformed into a triplet, and back and forth, but it is the underlying coherence and not the population exchange that seems to be directly {\it correlated} with the figure of merit $\delta Y_{\rm S}$. {\it It should be clear that we were able to arrive at these conclusions because we have an explicit quantifier of S-T coherence}.

To be precise, however, we should limit the affirmative answer to this statement: what the previous findings demonstrate is that S-T coherence {\it allows} for a quantum advantage not present in conditions of singlet-triplet incoherence. Whether such advantage is actually realized in nature, or in other words, whether the actual molecular parameters of the naturally occuring compass are such that the compass operating point is among those exhibiting large S-T coherence is a different question. However, this is not of fundamental interest for quantum biology. Of interest is what is in principle possible with such biochemical spin-dependent reactions. If it is found that they naturally work in a regime of large S-T coherence it would be quite an exciting finding, but even if this is not the case, knowing what is in principle possible would allow for the design of an artificial compass (or magnetometer in general) taking advantage of quantum coherence effects.

Put differently, since the correlation coefficient of $\delta Y_{\rm S}$ with $\overline{\cal C}$ is at the level of 0.3 (Fig. \ref{corr}c), the mean of $\delta Y_{\rm S}$ over the 5000 points is seen to drop with the decreasing mean of $\overline{\cal C}$ (Fig. \ref{corr}b), but not excessively, i.e. by about 60\% from $K_d=0$ up to $K_d=10k$. In contrast, considering the subset of highest $\delta Y_{\rm S}$ for each $K_d$, the drop with decreasing $\overline{\cal C}$ is much more significant (order of magnitude). However, as previously mentioned, it is unknown if the natural compass has evolved exploring those molecular parameters placing it at the high-$\delta Y_{\rm S}$ and high-$\overline{\cal C}$ regime (the upper part of the stripes in Fig. \ref{corr}a). 
\subsection{The role of the exchange interaction}
\begin{figure}[th!]
\begin{center}
\includegraphics[width=8 cm]{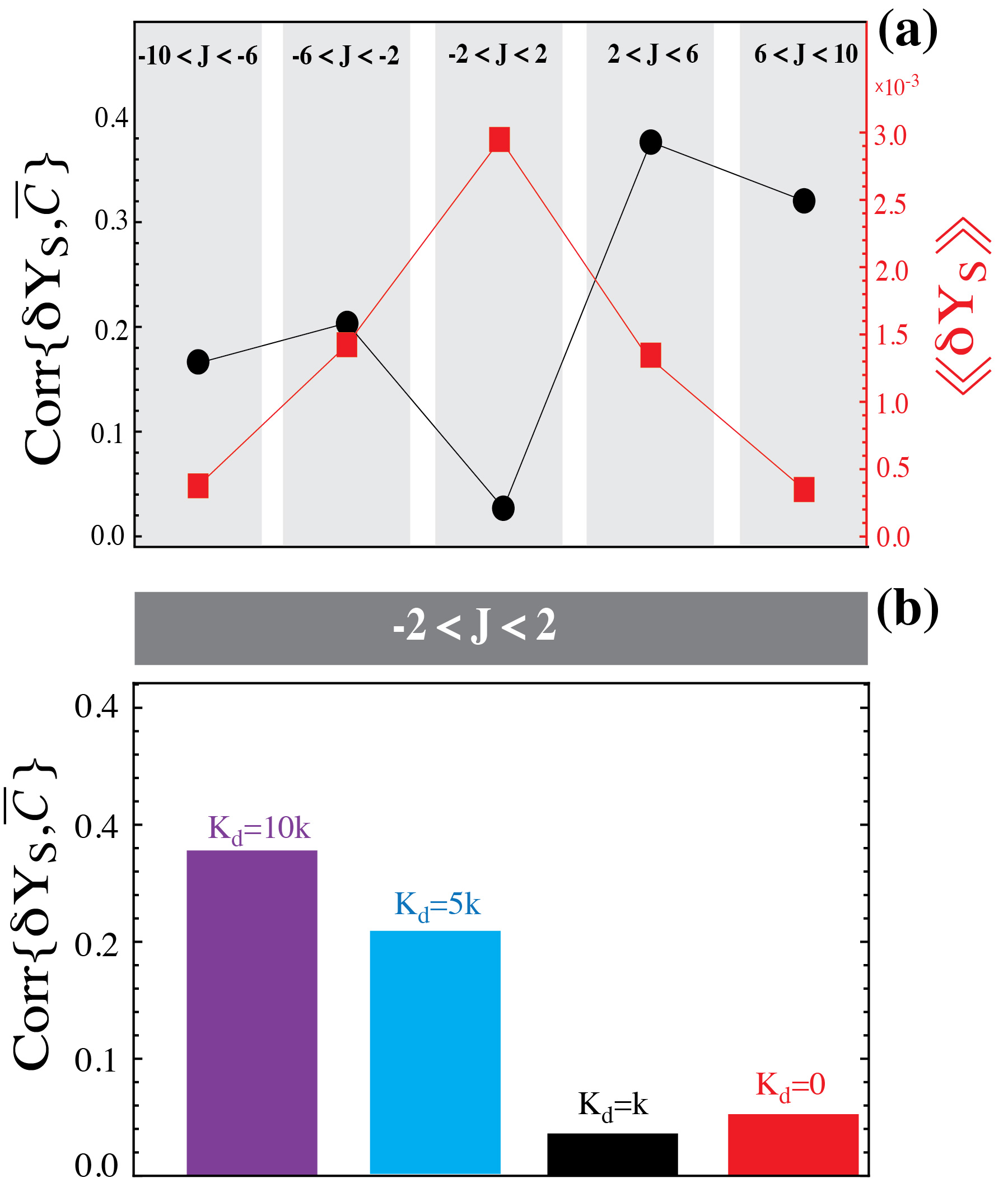}
\caption{(a) Correlation between the compass figure of merit $\delta Y_{\rm S}$ and the singlet-triplet coherence measure $\bar{\cal C}$ (left y-axis, black discs) and the mean value of $\delta Y_{\rm S}$ (right y-axis, red squares), for various subintervals of the exchange coupling, shown in the grey boxes (in units of the reaction rate $k$). Here $K_d/k=1$. (b) Same correlation, but for the particular sub-interval $|J|<2$ and for all four values of $K_d$.}
\label{Jcorr}
\end{center}
\end{figure}
The exchange interactions are known to play a subtle role in magnetoreception \cite{Efimova,Dellis}. To explore their role, we included them in the Hamiltonian with an exchange coupling in the interval $-10k\leq J\leq 10k$. Now we split this interval into 5 sub-intervals of width $\Delta J=4k$, and study the correlation of the figure of merit $\delta Y_{\rm S}$ with the S-T coherence $\bar{\cal C}$. The result is plotted in Fig. \ref{Jcorr}a for the case $K_d=k$. In the same figure (right y-axis) we also plot the mean vaule, $\langle\langle\delta Y_{\rm S}\rangle\rangle$, of $\delta Y_{\rm S}$ for those same subintervals of $J$. 

The observed behavior of $\langle\langle\delta Y_{\rm S}\rangle\rangle$ with $J$ is known since the work of \cite{Efimova}, i.e. it is already known that large values of $J$ supress the figure of merit $\delta Y_{\rm S}$ of the compass, as is also evident in Fig. \ref{Jcorr}a. We here observe two additional, counterintuitive effects. Namely, (i) for small values of $J$ it is the {\it correlation} between singlet-triplet coherence $\bar{\cal C}$ and the figure of merit $\delta Y_{\rm S}$ that is significantly suprressed, {\it to recover for large values of} $|J|$. (ii) There is also an asymmetry in this correlation between $J>0$ and $J<0$. 

The interpretation of both of these observations, being rather challenging, will be left as an open problem. However, we will speculate on the interpretation of Fig. \ref{Jcorr}. If the exchange coupling $|J|$ for a realistic compass in indeed large, {\it we can still make the case} that S-T coherence is a resource {\it because of the large correlation} observed between $\delta Y_{\rm S}$ and $\bar{\cal C}$ at large $|J|$, albeit at smaller absolute values of $\delta Y_{\rm S}$. On the other hand, if $|J|$ is small, in Fig. \ref{corr}b we observe that the correlation between $\delta Y_{\rm S}$ and $\bar{\cal C}$ recovers for large values of $K_d$. Based on this, we anticipate that there is another regime of the reaction dynamics, which we have not addressed in this work, and where the correlation could be significant even for small $|J|$. Namely, we here considered equal recombination rates $\ks=\kt=k$, in order to simplify reaction dynamics and decouple from the ongoing discussion on the form of the reaction super-operators. However, the regime $\kt\gg\ks$ is also interesting \cite{Dellis}, and based on Fig. \ref{corr}b it is conceivable that in this regime there is a large correlation between figure of merit and S-T coherence for all values of the exchange coupling $J$.
\section{Conclusions}
Quantum coherence is a fundamental resource for modern quantum technology. Its formal quantification has been established in recent years. One quantifier of quantum coherence is the quantum relative entropy between the density matrix describing the quantum state of the system under consideration and its diagonal version. 

We have here adapted this quantifier to radical-pairs, defining singlet-triplet coherence as the relative entropy between the radical-pair density matrix and its {\it block-diagonal version in the singlet-triplet subspaces}. We have then established the properties of this singlet-triplet coherence quantifier at a formal level. Having an explicit quantifier of singlet-triplet coherence, one can study the fundamental properties of biological magnetic sensing in various regimes, for example in regimes where the relevant spin states are highly coherent or highly incoherent. 

By doing so, we have shown that singlet-triplet coherence is indeed a quantum resource for magnetoreception, since (i) it is highly correlated with the figure of merit of the radical-pair compass, (ii) {\it both} singlet-triplet coherence {\it and} the figure of merit decrease significantly in the presence of singlet-triplet dephasing, and (iii) the singlet/triplet populations remain on average unaffected by such dephasing. 

Finally we explored the subtle role played by exchange interactions in promoting the correlation between singlet-triplet coherence and the figure of merit of the chemical compass. Along the same lines one could explore the role of other interactions entering the Hamiltonian, in particular when more nuclear spins are included. 

Last but not least, while defining the incoherent operations needed in the formulation of singlet-triplet coherence quantification, we gave an example where incoherent operations {\it on nuclear spins only} can have a significant effect on the singlet-triplet coherence, which is of electronic nature. This can happen when nuclear spins and electrons in the radical-pair are entangled, which is in general the case. This observation opens up a promising direction of studying the effects of nuclear spin dynamics, e.g. the interaction with the environment of the radical-pair's nuclear spins, and their consequences on radical-pair spin dynamics.


\begin{thebibliography}{10}
\bibitem{note}
We note that the dephasing term $-k(\qs R\qt+\qt R\qs)$ is of a Lindblad form, because due to the identity $R=\qs R\qs+\qt R\qt+\qs R\qt+\qt R\qs$, the dephasing term is identically equal to $-k(\qs R+R\qs-2\qs R\qs)$, which is a form frequently used in our previous work.
\end{thebibliography}
\end{document}